\documentclass[runningheads]{llncs}
\usepackage[utf8]{inputenc}
\usepackage[T1]{fontenc}
\usepackage{graphicx}
\usepackage{todonotes}

\begin{document}
\title{Explainable Point-Based Document Visualizations}
\author{Primož Godec\inst{1,2} \and Nikola Đukić\inst{1}\and Ajda Pretnar\inst{1} \and Vesna Tanko\inst{1,2} \and Lan Žagar\inst{1,2} \and Blaž Zupan\inst{1}}
\authorrunning{P. Godec et al.}
\institute{University of Ljubljana, Faculty of Comp. and Info. Science, Ljubljana, Slovenia  \and Revelo, Information Technologies d.o.o, Ljubljana, Slovenia}
\maketitle              
\begin{abstract}
Two-dimensional data maps can visually reveal information about the relations between data instances. Popular techniques to construct data maps are t-SNE and UMAP. The resulting point-based visualizations, though, provide information only through their interpretation. We here consider a set of abstracts from the articles on longevity to argue for using keyword extraction methods to label clusters of documents in the map. Among the considered approaches, the best results were obtained by recently proposed YAKE!. Surprisingly, a classical TF-IDF term ranking outperformed graph and embedding-based techniques.

\end{abstract}
\section{Introduction}

Science, according to Wikipedia, is ``a systematic enterprise that builds and organizes knowledge''. One of the earliest means of information organization is a map. Today, an important subfield of data science strives to map multi-dimensional data onto a two-dimensional plane to expose the relations between data instances. Prominent examples of such mappings include multi-dimensional scaling, t-SNE, and UMAP. These techniques embed data instances into point-based visualizations by preserving the distances (MDS) or neighborhoods (t-SNE, UMAP). The constructed visualization can be regarded as a map. Map's actual value, though, is obtained through explanations, where we associate regions or clusters of points with labels that expose their shared properties. The labels are essential to interpreting data maps; they provide means of analyzing map neighborhoods and differentiating distinct regions. 

In the manuscript, we focus on constructing data maps from the corpus of textual documents. We study keyword extraction mechanisms to label clusters of textual documents. Keyword extraction is a task where keywords or key phrases are selected among words and phrases in the document~\cite{campos2020}. Several techniques for keyword extraction have been proposed. For our purpose, we focus on {\em unsupervised} approaches, which are domain-independent and do not need a labeled training data set. Among these, the TF-IDF~\cite{jones1972} is a baseline approach that estimates word importance in the document relative to the entire corpus. It computes the frequency of the term in the document weighted by the inverse of the frequency in the whole corpus. A more advanced approach is YAKE!~\cite{campos2020} which uses statistical features such as the position and frequency of words, context information, and the term's spread in the document. In contrast to TF-IDF, it extracts keywords on a single document basis and does not need a large corpus. Graph-based methods develop a graph of related phrases from the documents and use graph ranking methods to score the phrases. RAKE~\cite{rose2010} builds a graph of term concurrences and uses term frequency, term degree in the graph, or a combination of both as a score. It also combines term scores into phrase scores. Deep learning has enabled embedding-based approaches. These allow embedding candidate phrases and documents into the same vector space and assigning phrases with the highest similarity to the respective documents~\cite{bennani2018}.

We here compare the keyword extraction approaches we have reviewed above and then use the best-performing method on a corpus of paper abstracts from the domain of longevity. A subset of abstracts included a list of keywords provided by authors that we have used to assess recall and precision of the examined approaches. We then show that the keyword extraction approach assigns meaningful labels to a map of longevity papers and thus provides means for explaining and interpreting the document map.

\section{Methods and Data}

Our input is a set of unstructured textual documents that, in principle, does not contain any labels. We propose to construct a two-dimensional map and characterize its clusters through keyword extraction. The proposed analysis pipeline borrows from Han {\em et al.}~\cite{han2018}, who used TF-IDF to infer cluster labels.

\vspace*{2mm}\noindent{\bf Document embedding.} Our first step towards explainable document visualizations is to vectorize the text documents and embed them into familiar and understandable two-dimensional point-based visualizations. We use fastText~\cite{grave2018} for vectorization of words and compute the whole-document vector profile as the weighted average over all words in the document. The high-dimensional vector representations of all documents are then embedded into a two-dimensional space by t-SNE~\cite{laurens2008}. The resulting visualization shows a set of documents represented by points and can already uncover similarities between groups of documents.

\vspace*{2mm}\noindent{\bf Clustering.} We fit a Gaussian mixture model with a predefined number of components to mapped data. Documents (points) are assigned to the most likely gaussian component, or they can not be associated with any component if the probability of membership is below a threshold (we used $P<=0.6$). Notice that alternatively, we could use some other clustering technique, like $k$-means clustering or DBSCAN, where the first one suffers from the constrained shape of clusters, and the second from high sensitivity to parameters of the method.

\vspace*{2mm}\noindent{\bf Keyword extraction.} We characterize the identified clusters with representative keywords. These are first extracted from each document using one of several possible methods. Our candidate methods for keyword extraction included YAKE!, RAKE, TF-IDF, TextRank, and Embedding. We represent each document with the $k=20$ best-scored keywords. We then check which keywords appear statistically more frequently in the cluster of documents compared to the expected occurrences across all documents for each cluster. According to the cluster's frequency, we first score cluster-specific terms and then select them for labeling if they are cluster-enriched so that their $p$-value according to hypergeometric test, comparing their frequency within and outside the cluster, is lower than $0.05$.

\vspace*{2mm}\noindent{\bf Data.} We have tested the keyword extraction techniques and assessed the proposed document mapping and labeling approach on the articles from the topic of {\em longevity}. We downloaded 16,791 abstracts with the longevity MeSH Term from PubMed (\url{https://pubmed.ncbi.nlm.nih.gov}). Author-assigned keywords were a\-vailable for a subset of 4,594 articles, where, on average, a single article was assigned 6.2 keywords. We have used the keywords to objectively compare keyword extracting methods and assess recall and precision. We used the best-scored keyword extraction technique to infer cluster labels in the two-dimensional map, aiming at showing -- in a intuitive way -- that they can help in interpretation.

\section{Experiments \& Results}

We ran two experiments. The aim of the first one was to assess the quality of various keyword extraction approaches. We have purposely selected different approaches that range from the simplest TF-IDF to those using embedding by deep-learning. The comparison of these methods was quantitative and based on the accuracy of retrieval of author-assigned keywords to article abstracts. In the second experiment, we have employed best-performing approach to label clusters in the map of documents. The evaluation of the quality of the map was subjective and based on interpretability of the results.

\begin{figure}[htb]
	\centering
 	\includegraphics[width=0.9\textwidth]{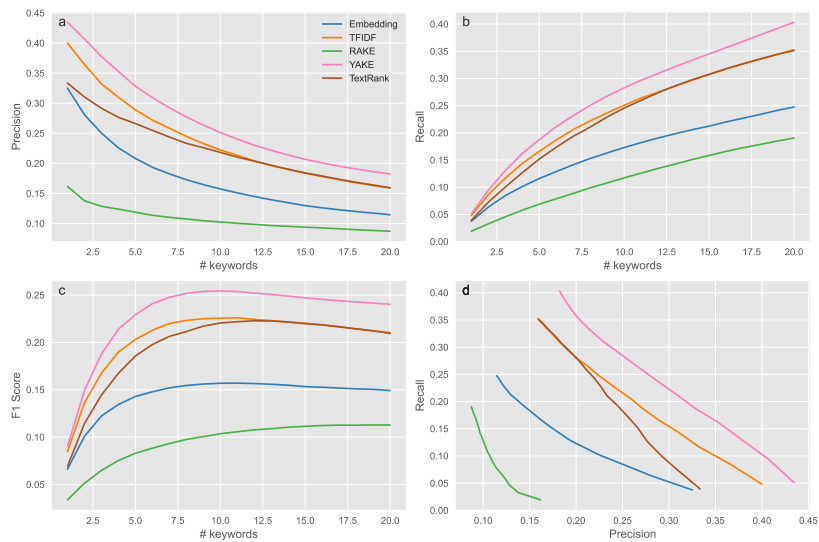}
	\vspace{-2mm}
 	\caption{The graphs show (a) the average precision for each method for different numbers of selected best-scoring keywords, (b) the recall for each method, (c) the F1 score, and (d) a combination of precision and recall. In (a) to (c), the x-axis depicts the number of terms considered as keywords.}
 	\label{fig:method-compare}
 	\vspace{-5mm}
\end{figure}

\vspace*{2mm}\noindent{\bf Comparison of Keyword Extraction Methods.} Our aim here is to find the optimal method for keyword extraction from the domain-specific corpus. We infer keywords for each document with each of the keyword-extraction techniques and compare them with the authors' keywords. \figurename~\ref{fig:method-compare} shows average precision, recall, and F1 score where we have included a different number of best-scored terms in the keyword set. With more lower-ranked keywords included, the precision of the algorithms drops but recall increases. In the combined graph (d), the winning method with the highest recall and precision has the curve closest to the graph's top-right corner. The results are consistent with the F1 score, where YAKE! always wins for any number of considered keywords. Considering the results, the best performing method is YAKE!. Surprisingly, despite its simplicity, the second-best technique is TF-IDF, which performs better than more sophisticated graph-based and embedding techniques.

\begin{figure}[htb]
	\centering
 	\includegraphics[width=0.6\textwidth]{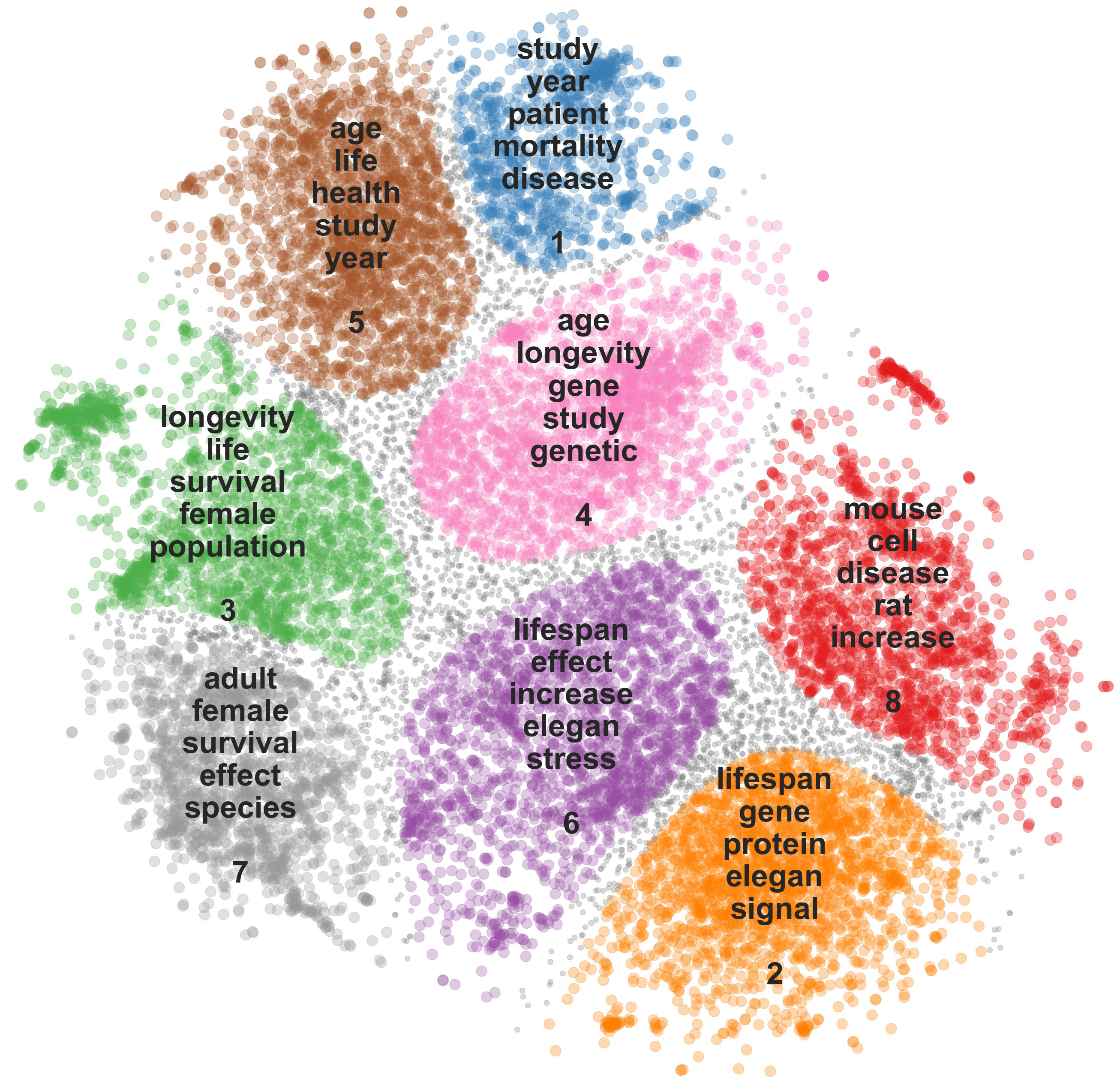}
 	\vspace{-2mm}
 	\caption{An annotated map of 16,791 articles on longevity.}
 	\label{fig:document-map}
 	\vspace{-5mm}
\end{figure}

\vspace*{2mm}\noindent{\bf Explainable Document Map.}
When dealing with corpora of hundreds or thousands of documents, it is challenging to observe keywords for each document separately. To get a broader image, we use embedding methods such as t-SNE to get a map of documents. We then infer groups of documents on the map and describe them with keywords. In our case, in \figurename~\ref{fig:document-map} we inferred the map of all 16,791 documents and annotated clusters with five enriched keywords.

We found cluster labels meaningful and representative. For example, in cluster 5 we found general articles that study longevity and where central documents, according to the Silhouette score, include titles such as ``Population aging and public health-the active aging concept'', ``Who were old in the Middle Ages'', and ``Longevity according to life histories of the oldest-old''. Differently, cluster 8 includes reports on model-based studies, with cluster-central documents bearing the titles such as ``Rodent carcinogenicity with the thiazolidinedione antidiabetic agent troglitazone'' and ``Prolonged lifespan and high incidence of neoplasms in NZB/NZW mice treated with hydrocortisone sodium succinate''. According to our intuition, other cluster labels equally well represented the topics of the documents in the clusters, supporting the value of our approach to explanation.

\section{Conclusion}

While embedding data sets in two-dimensional visualizations is standard, various keyword extraction approaches to characterize clusters of textual documents have not been studied extensively. We show that the combination of document embedding and label-based characterization of clusters can yield helpful and explainable document maps. Cluster labels hint to the content of documents in the cluster and can provide guidance for further exploration. Our experimental study includes only one specific data set. While comparison of keyword extraction methods would require a more elaborate study with various data sets, it is easy to envisage and devise other valuable examples of explainable document maps in the domain of medical texts.\footnote{The work was supported by a grant on semantic analysis of text from Slovenian Ministry of Public Administration.} 

\bibliographystyle{splncs04}
\bibliography{main.bib}
\end{document}